\documentclass{emulateapj}

\newcommand{\sigva}[0]{\left\langle \sigma_{\text{A}}v\right\rangle}

\usepackage{dcolumn}    
\usepackage{subfigure}  
\usepackage{amsmath}
\usepackage{graphicx}
\usepackage{times}

\begin{document}
\preprint{\dots}
\title{Searching for Dark Matter in Messier 33}
\author{Enrico~Borriello$^{1}$\footnote{To whom correspondence should be
addressed;\\E-mail:~eborriello@na.infn.it.},
Giuseppe~Longo$^{1,2}$, Gennaro~Miele$^{1,3}$, Maurizio~Paolillo
$^{1,2}$, Beatriz~B.~Siffert$^{1}$, Fatemeh~S.~Tabatabaei$^{4}$,
Rainer~Beck$^{4}$}

\affiliation{$^{1}$Dipartimento di Scienze Fisiche, Universit\`a
di Napoli "Federico II", Complesso
Universitario di Monte S. Angelo V. Cinthia, 9, I-80126, Napoli, Italy\\
$^{2}$INAF-OACN, via Moiariello 16, I-80128 Napoli, Italy\\
$^{3}$ INFN - Sezione di Napoli, Complesso Universitario di Monte
S. Angelo V. Cinthia, 9, I-80126, Napoli, Italy\\
$^{4}$ Max-Planck Institut f\"ur Radioastronomie, Auf dem H\"ugel
69, 53121 Bonn, Germany}

\date{\today}

\begin{abstract}
Among various approaches for indirect detection of dark matter,
synchrotron emission due to secondary electrons/positrons produced
in galactic WIMPs annihilation is raising an increasing interest.
In this paper we propose a new method to derive bounds in the
$m_\chi$--$\sigva$ plane by using radio continuum observations of
Messier 33, paying particular attention to a low emitting {\it
Radio Cavity}. The comparison of the expected radio emission due
to the galactic dark matter distribution with the observed one
provides bounds which are comparable to those obtained from a
similar analysis of the Milky Way. Remarkably, the present results
are simply based on archival data and thus largely improvable by
means of specifically tailored observations. The potentiality of
the method compared with more standard searches is discussed by
considering the optimistic situation of a vanishing flux (within
the experimental sensitivity) measured inside the cavity by a high
resolution radio telescope like ALMA. Under the best conditions
our technique is able to produce bounds which are comparable to
the ones expected after five years of Fermi LAT data taking for
an hadronic annihilation channel. Furthermore, it allows to test the hypothesis that space
telescopes like Pamela and Fermi LAT are actually observing
electrons and positrons due to galactic dark matter annihilation into leptons.
\end{abstract}

\maketitle

\section{Introduction}

Weakly Interacting Massive Particles (WIMPs) still represent the
most natural candidate for Dark Matter (DM) whose detection is one
of the main purposes of astrophysical and cosmological
observations nowadays (\citealp{Bertone:2004pz},
\citealp{Komatsu:2008hk}).

From a theoretical point of view, models of WIMPs arise
straightforwardly in supersymmetric extensions of the Electroweak
Standard Model as the lightest supersymmetric particle. These
candidates can annihilate in couples to produce ordinary particles
as final states  which, in principle, can be detected.

Concerning the produced electromagnetic radiation, gamma-rays
provide the most promising opportunity due to the very low
attenuation in the interstellar medium, and to its high detection
efficiency (see for example \citealp{Jungman:1995df},
\citealp{Bergstrom:2000pn}, \citealp{Bertone:2004pz} for a review
of this extensively studied issue). In this respect the activity
of the recently launched Fermi Large Area Telescope provides an
exciting chance to have a large amount of data on the gamma-ray
sky between 20 MeV and 300 GeV, including the largely unexplored
energy window above 10 GeV \cite{Collaboration:2009zk}.

Nevertheless, high energy electrons and positrons, once produced
in the annihilation cascade of WIMP pairs, interact with gas,
radiation and the magnetic field in the galaxy
(\citealp{Baltz:1998xv}, \citealp{Donato:2003xg},
\citealp{Hooper:2004bq}). During their propagation in the galactic
medium they release secondary radiation in the radio and X-ray
bands (\citealp{Blasi:2002ct}, \citealp{Tasitsiomi:2003vw},
\citealp{Aloisio:2004hy}, \citealp{Baltz:2004bb},
\citealp{Colafrancesco:2005ji}, \citealp{Colafrancesco:2006he},
\citealp{Bergstrom:2006ny}, \citealp{Regis:2008ij},
\citealp{Jeltema:2008ax}, \citealp{Zhang:2008rs}), giving rise to
a further signal of the DM annihilation that is, in principle,
detectable.

\begin{figure*}
\begin{center}
\includegraphics[width=7 cm, height=8 cm]{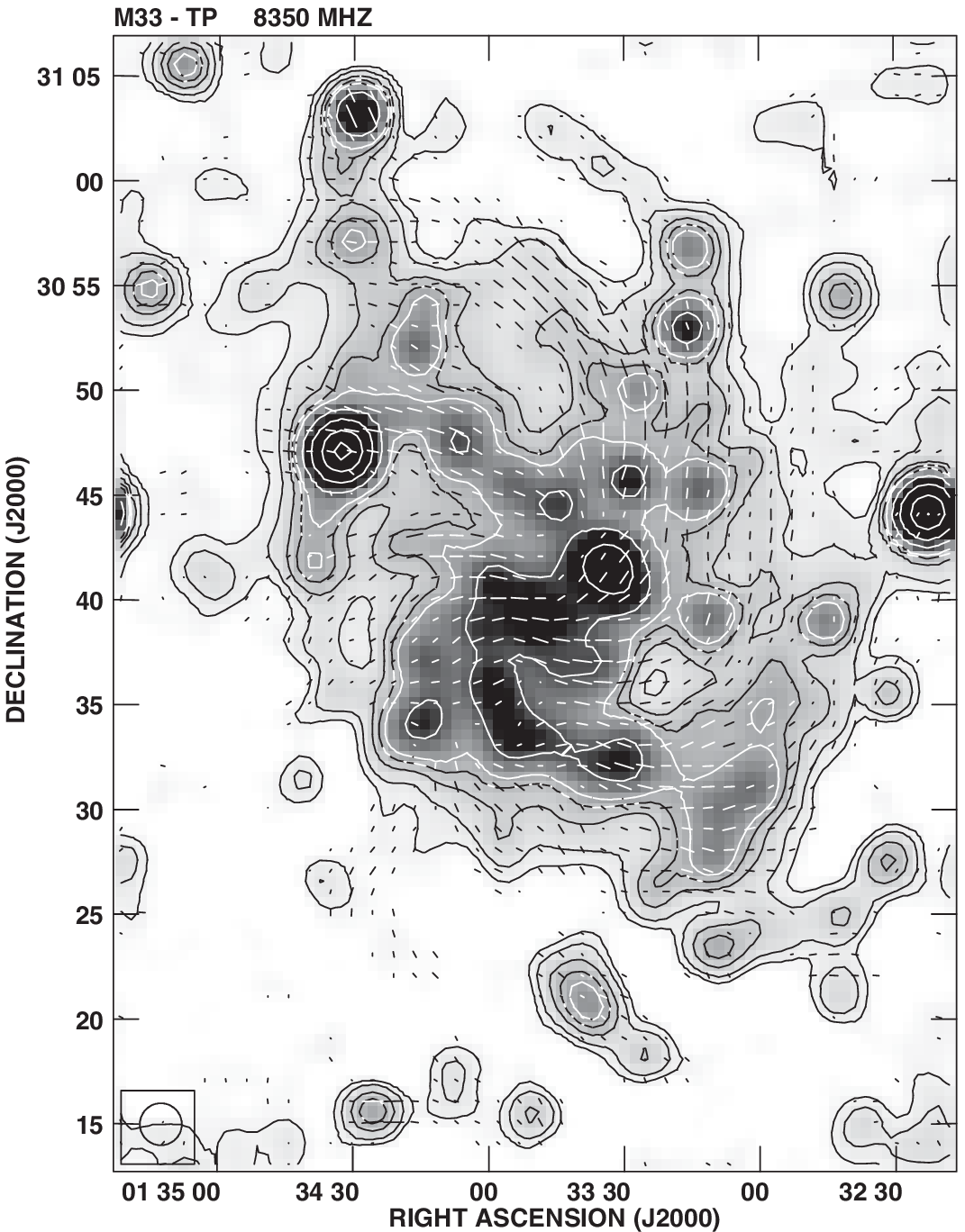} \label{Map3_6cm} \qquad
\includegraphics[width=7 cm, height=8 cm]{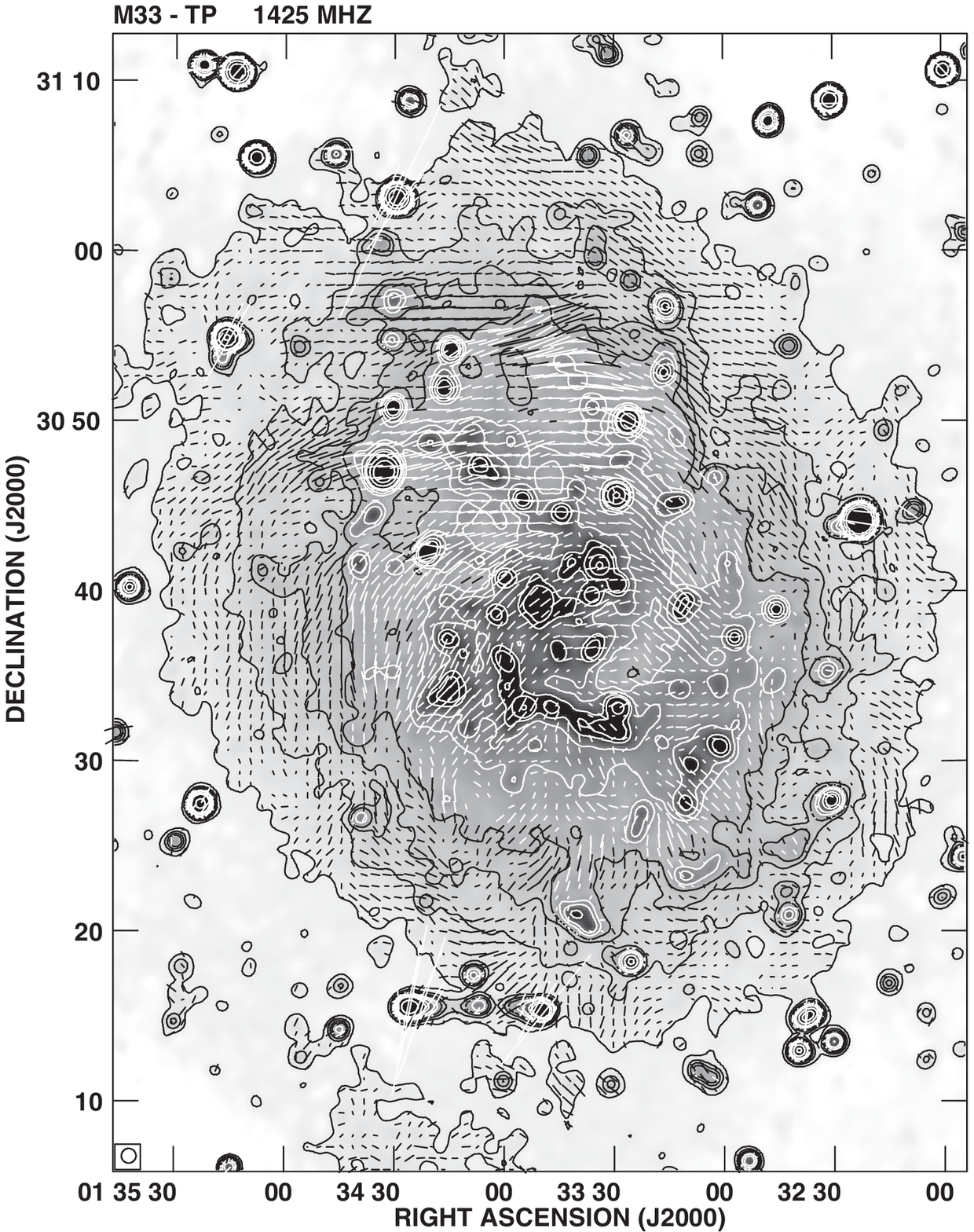} \label{Map20cm}
\caption{Radio continuum emission from M33 at $\lambda = 3.6$ cm
(left) and $\lambda = 20$ cm (right), with apparent {\bf
B}-vectors of polarized intensity superimposed. The beam areas are
shown in the left corner of each figure. Contour levels are 0.5,
1, 2, 3, 4, 8, 16, 32 and 64 mJy/beam for the $3.6$ cm and 0.3,
0.6, 0.9, 1.2, 1.8, 2.4, 3.6, 4.8, 9.6 and 19.2 mJy/beam at $20$
cm \cite{Tabatabaei2007a}.} \label{radioplots}
\end{center}
\end{figure*}

In \citealp{Borriello:2009gy}, some of the authors of the present
paper have developed a semianalytical approach to compute the
synchrotron emission from the Milky Way (MW) WIMP pairs
annihilation. The comparison of the theoretical DM radio signal
from our Galaxy with the observed emission, (where only the CMB is
modelled and removed) allows to draw an exclusion plot in the
$m_\chi$--$\sigva$ plane. For this purpose one can use the code
described in \citealp{deOliveiraCosta:2008pb} where most of the
whole sky radio observations in the range 10 MHz-100 GHz are
collected and a scheme is described to derive interpolated, CMB
cleaned sky maps at any frequency in this range. This result,
however, is strongly affected by two problems:
\begin{itemize}
\item[i)]
The difficulties in removing the foreground contribution from the
radio continuum. Such a removal would allow, at least, an
important refinement of the $m_\chi$--$\sigva$ bounds or, even
better, the detection of DM. Let us recall that different
approaches to the foreground removal at the WMAP frequencies
produced very different results, exhibiting a residual signal
(which has been interpreted as a signature of the DM annihilation)
as in \citealp{Finkbeiner:2004us} and \citealp{Hooper:2007kb} or
no additional feature, as in a more recent analysis performed by
the WMAP team itself \cite{Gold:2008kp}.

\item[ii)]
The fact that the measured quantities are averaged by the
integration along the line of sight which prevents a proper
estimate of the local values of both galactic magnetic field and
radio continuum emission. The high level of uncertainty on the
local values of the MW  magnetic field propagates on the results
after the integration along the line of sight (an order ten kpc
long path) is performed.

\end{itemize}

To overcome problems ii) we will reapply such an analysis to the
nearby, close to face--on, $Scd$ galaxy Messier 33 (M33), taking
advantage of the thinness of the region source of the signal. This
galaxy is the third largest in the Local Group and is located at a
distance of $840$ kpc from us \cite{Freedman1991}. It has been
extensively studied at all wavelengths: radio continuum
(\citealp{Tabatabaei2007a}, \citealp{Tabatabaei2007b},
\citealp{Tabatabaei2008}), far infrared \cite{Tabatabaei2007c},
near IR \cite{Cioni:2007}, optical \cite{Massey:2006}, UV
(\citealp{Ciani:1984}, \citealp{Gordon:1999}), and X--ray
\cite{Haberl:2001}. Its kinematics has also been the subject of
many studies \cite{Carignan:2006}. M33 has an inclination of $\sim
56^{\circ}$ and due to its rather flocculent nature, it presents
well defined arm and interarm regions. Some of the latter regions
can be characterized by a particularly low level of radio
emissivity (hereafter denoted for simplicity {\it Radio
Cavities}), even though placed not far from the galactic center,
and thus where possibly a non negligible magnetic field is at
work. Hence, the poor radio emission would be simply due to a lack
of astrophysical electrons/positrons. In this framework, the
presence of a halo of WIMPs would inject in any case a flux of
electron/positrons. These particles, while diffusing in the galaxy
and propagating in the magnetic field, have to produce a radio
signal compatible with the observed one. The requirement that the
DM induced radio signal coming from a particular {\it Radio
Cavity} of M33 does not overcome the observed one
straightforwardly fixes bounds in the $m_\chi$--$\sigva$ plane.

Analyses devoted to the study of radio emission by external
galaxies are already present in literature (see for example
\citealp{Tasitsiomi:2003vw} and \citealp{Colafrancesco:2006he}),
even if not focussing on {\it Radio Cavities}. It is worth
noticing that the main advantage of considering a small region
rather than the whole galaxy is that this approach allows to keep
the different astrophysical hypothesis much more under control.
For the purposes of the present work, M33 provides an ideal case
since it is near, it has a small bulge--to--disk ratio, and
therefore, it is possible to trace the inter--arm regions also
relatively close to the central and densest regions. Moreover, due
to its relatively small inclination (e.g. compared to M31), it
allows to minimize the effects of the integration of the signal
along the line of sight.

Unfortunately we cannot solve the first problem. The radio
emission of galaxies like M33 has three components: the clumpy
thin disk ($\sim200$ pc scale height) with thermal and synchrotron
emission from star-forming regions and supernova remnants, the
smooth thick disk ($\sim1$ kpc) with mostly synchrotron emission
by young electrons originating from the thin disk, and the smooth
halo (unknown extent) with synchrotron emission by old electrons
from the thin disk. The relative contributions may vary from
galaxy to galaxy, but in general the radio from the thick disk
dominates, as visible in edge-on galaxies \cite{Beck2009}. The
emission from the dark matter halo will mix with the thick disk
emission but the mixing ratio is of course unknown. For this
reason we preferred a conservative approach not subtracting any
possible astrophysical component.

The paper is organized as follows. In Section \ref{Sec:cavity} we
describe the main characteristics of M33 {\it Radio Cavity},
whereas in Section \ref{sec:DMsignal} we compute the expected
radio signal coming from it. Finally, in Sec.
\ref{Sec:conclusions} we derive our constraints for WIMPs in the
$m_\chi$--$\sigva$ plane and give our conclusions.

\section{The \textit{radio cavity} in M33}
\label{Sec:cavity}

In the present analysis, we use the observations of M33 at
$\lambda=3.6$ cm and $\lambda=20$ cm described in
\citealp{Tabatabaei2007a} and reported in Fig. \ref{radioplots}.

A detailed description of the data acquisition reduction and
analysis can be found in the above quoted reference. Here we shall
just point out a few relevant facts. The data at $\lambda = 3.6$
cm were obtained using the 100-m Effelsberg radiotelescope and the
final map has a HPBW of $80^{\prime\prime}$ and an r.m.s noise of
$\sim 220 \, \mu$Jy/beam. On the other hand the data at $\lambda =
20$ cm were obtained using the VLA-D array and corrected for lack
of flux, due to missing short spacing, using single dish 20\,cm
observations obtained at the Effelsberg radiotelescope. The
resulting map has an HPBW of $51^{\prime\prime}$ and an r.m.s
noise of $\sim 70 \,\mu$Jy/beam.

\begin{figure}[!t]
\begin{center}
\includegraphics[width=1.0\columnwidth]{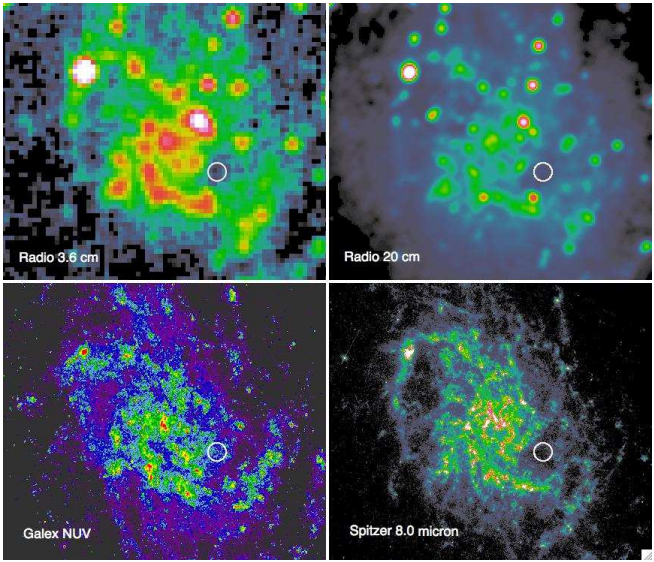}
\caption{Emission from M33 at different wavelengths. From the
upper left clockwise: radio continuum at 3.6 cm, radio continuum
at 20 cm, far infrared at 8.0 $\mu m$ (data from Spitzer
telescope), and Near UV at 2300 \AA. The {\it Radio Cavity} is
marked by a circle.} \label{M33_multiwav}
\end{center}
\end{figure}

For the purpose of this work, the most noticeable feature, marked
as a circle in Fig.~\ref{M33_multiwav}, is the interarm cavity
located at RA $\simeq 1^{{\rm h}}$33$^{{\rm m}}$24.0$^{{\rm s}}$
and Dec $\simeq 30^{\circ} \ 35^{\prime}\ 39.0^{\prime\prime}$,
i.e. at $\simeq 0.14^{\circ}$ SE of the bright nucleus (2.1 kpc
from the center). This cavity has an area of $\sim 5 \times
10^{3}$ arcsec$^{2}$, and is therefore well resolved at all
wavelengths.

Despite the small optical and radio flux coming from the cavity,
there is no reason to believe the magnetic field there to be
significantly smaller than the one in the surrounding regions.
Therefore, the cavity is the ideal place to look for a DM signal,
since there the ratio between the DM radio signal and the
astrophysical foreground reaches its maximum value.

\section{DM synchrotron signal}
\label{sec:DMsignal}

Let us first assume that M33 is perfectly face--on (later we will
correct for the angle it forms with the line of sight) and
consider a cartesian galactocentric coordinate system $(x,y,z)$
with the $z$--axis pointing in our direction. If $j_{\nu}$ is the
synchrotron emissivity due to DM annihilation, the radio flux
coming from an elementary volume $dV$ within M33 is given by
$dI_{\nu}=j_{\nu}dV/4\pi D^2$, where $D$ is the distance of M33
from us.

Thanks to the face--on approximation and the large value of $D$,
we can simply write $dV=D^2 d\Omega dz$ and consider the
integration along the line of sight as a simple integration along
the $z$--axis. At last, the thinness of the integration volume
allows us to take into account the angle $\theta$ the galaxy forms
with the line of sight by simply multiplying the flux by
$\cos{\theta}$. The result is
\begin{equation}
 \frac{dI_{\nu}}{d\Omega}(\alpha,\beta)=
 \frac{\cos{\theta}}{4\pi}\int j_{\nu}(x,y,z)dz \ ,
\label{eq:diff_flux}
\end{equation}
where $\alpha=(x/D)\cos\theta$ and $\beta=y/D$ are two visual angles
and where we chose the $y$--axis to be coincident with the M33 major axis.

We perform the integration of Eq. \ref{eq:diff_flux} over the
range $z \in [-2\,\textrm{kpc},2\,\textrm{kpc}]$. This choice does
not sensibly affect the results due to the exponential decrease of
the DM synchrotron emissivity away from the galactic disk.

An analytical expression of $j_{\nu}$ can be found in
\citealp{Borriello:2009gy} in terms of the DM density profile, the
magnetic field and the interstellar radiation field (ISRF). The
remaining parts of this section are therefore devoted to the
parametrization of the \textit{Radio Cavity} in M33.

\subsection{Dark Matter density profile}

It is not yet possible to uniquely determine the DM density of
M33, nonetheless both a cored and a spiked profile can be deduced
fitting the galaxy rotation curve (see Fig. 5 of
\citealp{Corbelli2003}). The cored case is well represented by a
Burkert \cite{Burkert1995} density profile
\[
\rho(r)=\left(1+\frac{r}{r_0}\right)^{-1}\left[1+\left(\frac{r}{r_0}\right)^2\right]^{-1}\rho_0 \ ,
\]
while the spiked profile is well described by and a NFW density
distribution \cite{Navarro:1996gj}
\[
\rho(r)=\left(\frac{r}{r_{0}}\right)^{-1}\left(1+\frac{r}{r_{0}}\right)^{-2}\rho_{0} \ .
\]
We consider two of the various fits obtained in \citealp{Corbelli2003}.
Their defining parameters are:
\begin{center}\begin{tabular}{|l|c|c|}\hline
model   & $r_0$ (kpc)   & $\rho_0$ (GeV\,$c^{-2}$\,cm$^{-3}$)   \\ \hline
NFW & 35        & 0.0574                \\
Burkert & 12        & 0.420                 \\ \hline
\end{tabular}\end{center}

Profiles with a central slope steeper than $r^{-1}$ like the Moore
profile \cite{Moore1999} are excluded \cite{Corbelli2003}.

\begin{figure*}[t]
\begin{center}
\includegraphics[width=.33\textwidth]{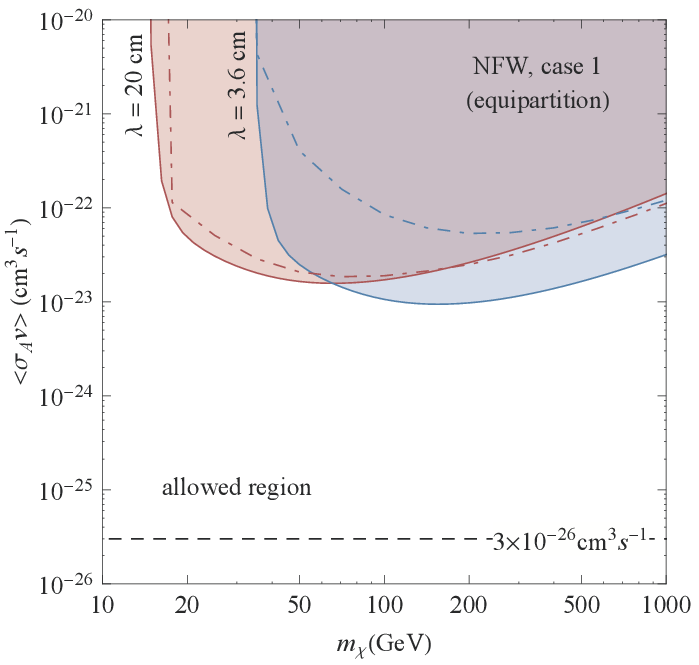}
\includegraphics[width=.33\textwidth]{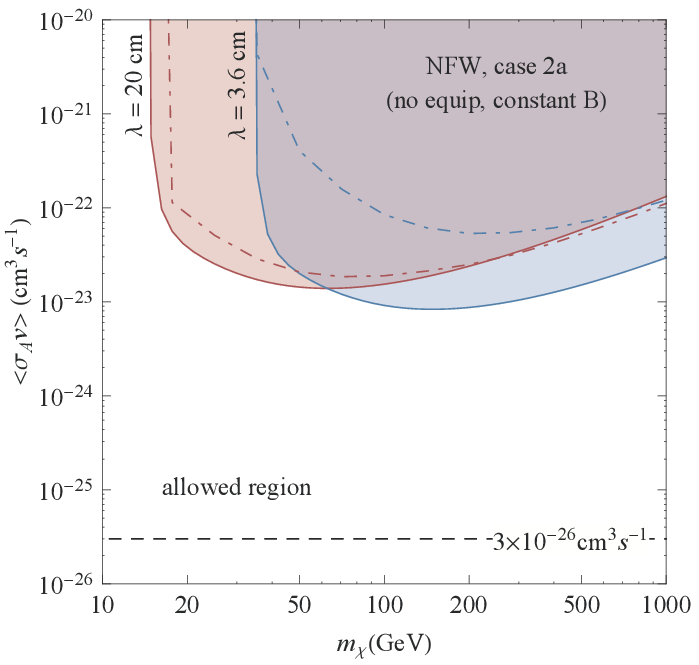}
\includegraphics[width=.33\textwidth]{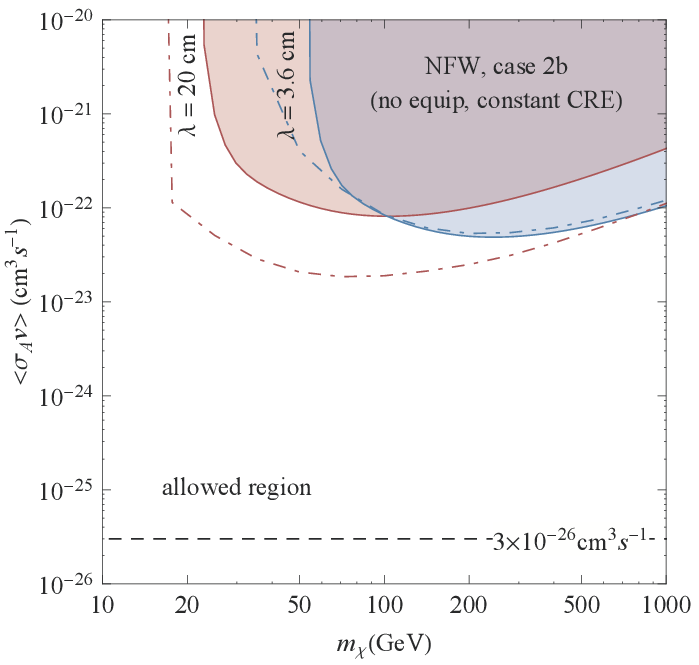}\\
\includegraphics[width=.33\textwidth]{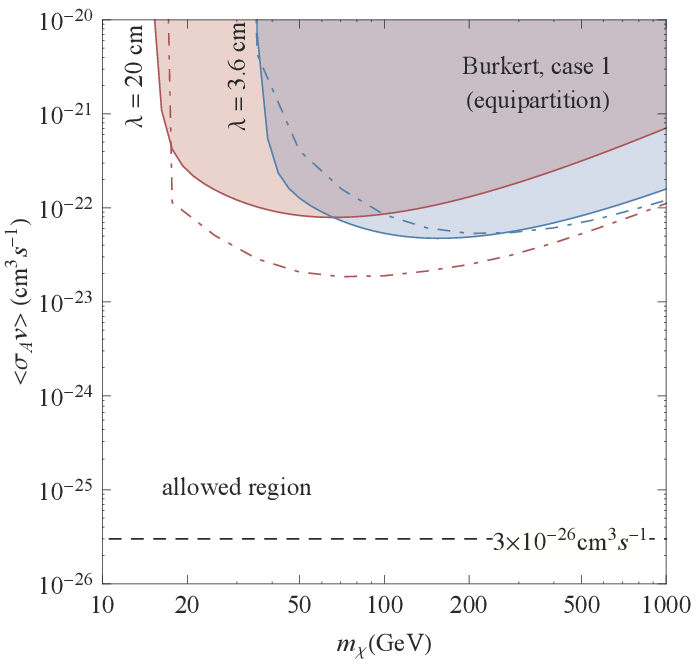}
\includegraphics[width=.33\textwidth]{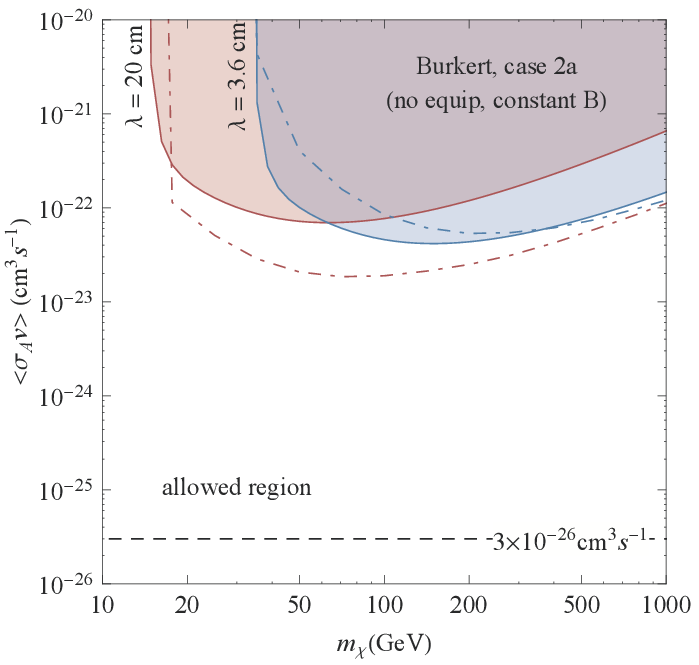}
\includegraphics[width=.33\textwidth]{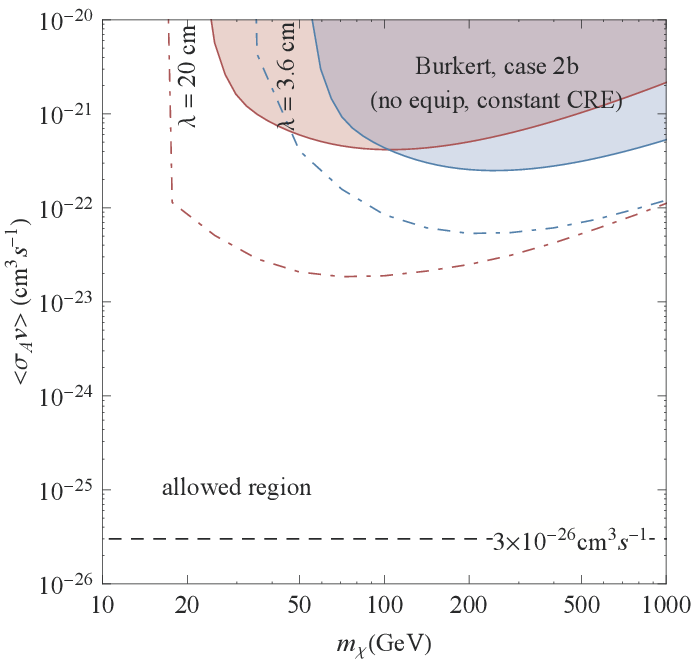}
\caption{Constraints in the $m_{\chi}$--$\left\langle \sigma_A v
\right\rangle$ plane from radio observation of the \textit{Radio
Cavity} of M33 at 3.6 and 20 cm. Full annihilation into $b\bar{b}$
couples is assumed. The upper panels refers to a NFW DM density
profile, the lower ones to a Burkert profile. Three cases are
considered: (1) the equipartition hypotesis is verified; (2a) it
is not and $B$ is constant across the cavity; (2b) it is not and
CRE density is constant across the cavity. Solid lines refer to
M33, while dot--dashed lines show the bounds obtainable -- under
the same physical hypothesis -- for the MW (as in
\cite{Borriello:2009gy}).} \label{hadronicbounds}
\end{center}
\end{figure*}
\begin{figure*}[t]
\begin{center}
\includegraphics[width=.33\textwidth]{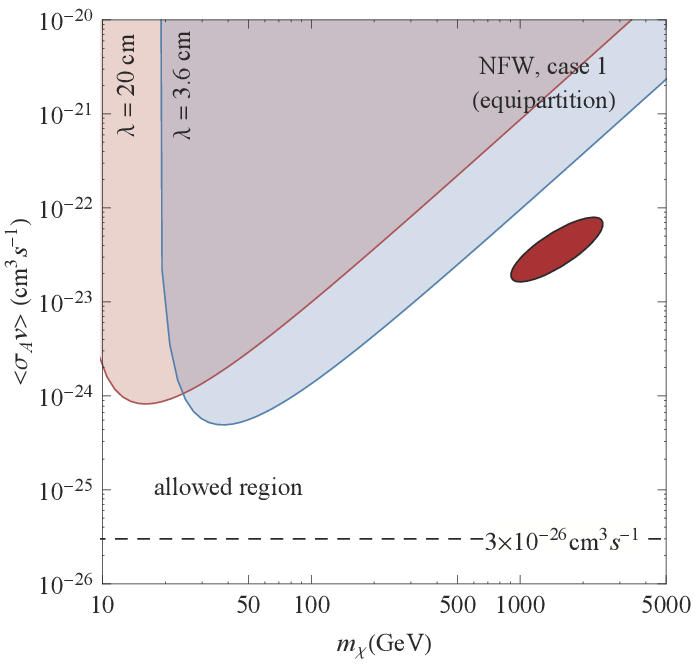}
\includegraphics[width=.33\textwidth]{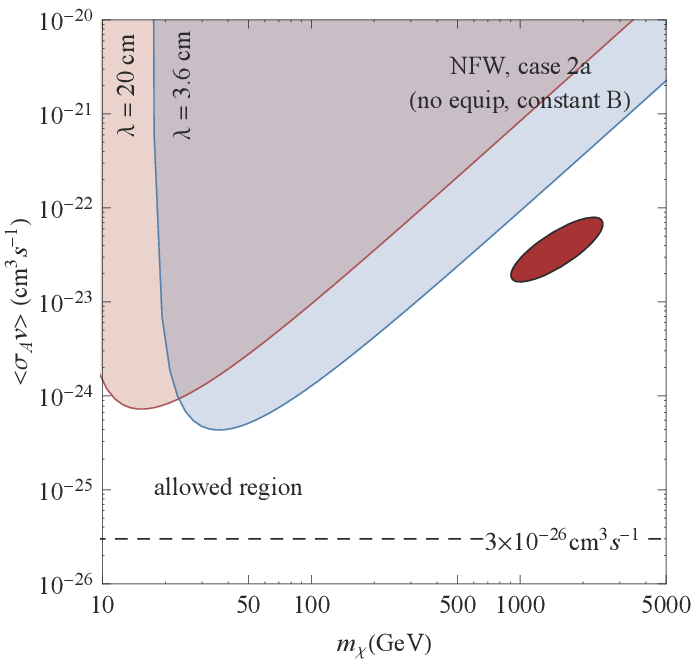}
\includegraphics[width=.33\textwidth]{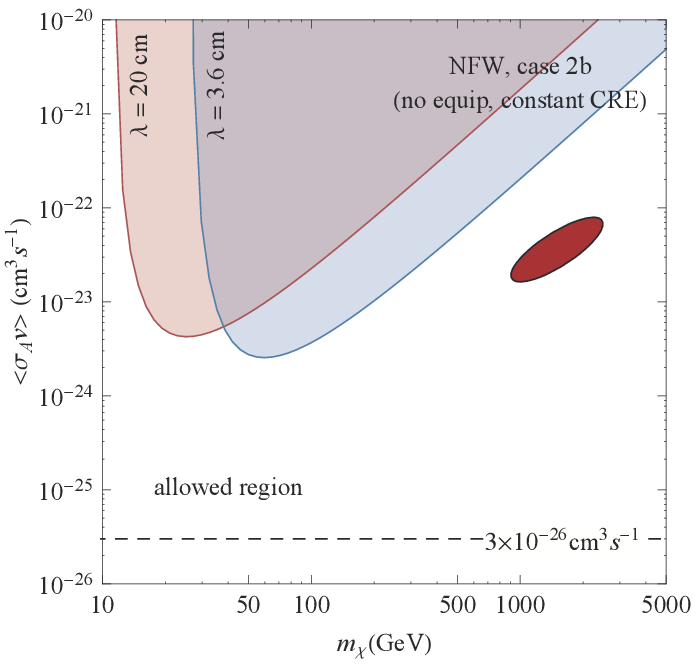}\\
\includegraphics[width=.33\textwidth]{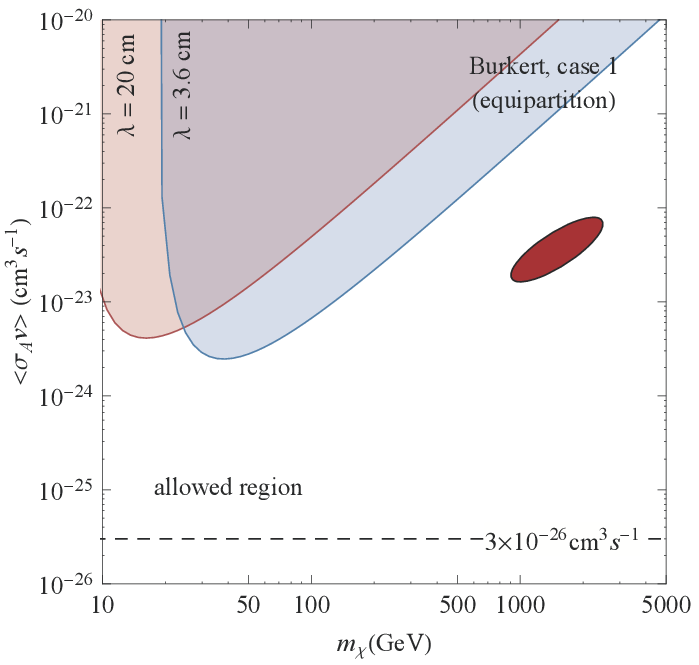}
\includegraphics[width=.33\textwidth]{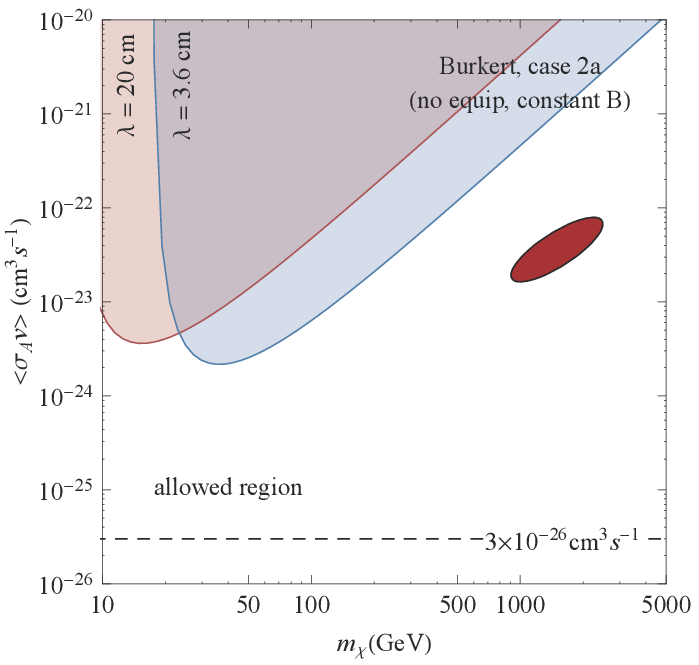}
\includegraphics[width=.33\textwidth]{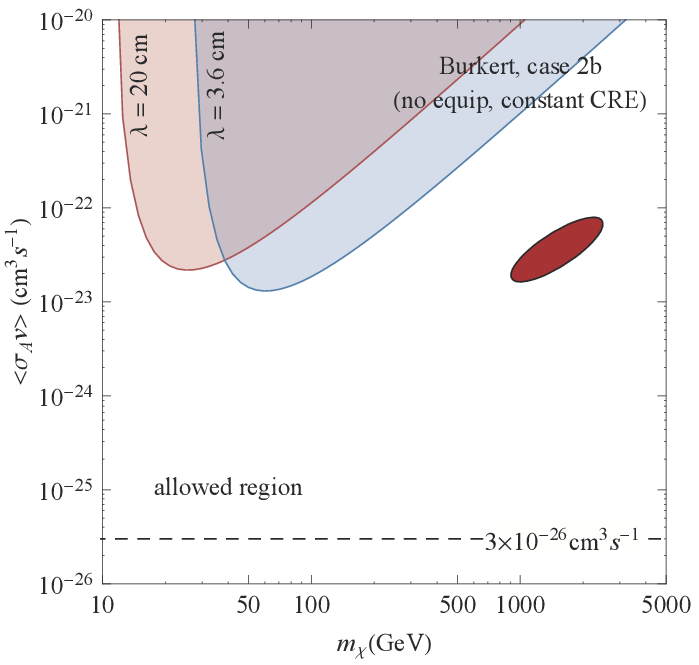}
\caption{As in Fig. \ref{hadronicbounds} but for DM particles
fully annihilating into $\mu^+\mu^-$ couples. This time the
benchmark case is represented by the Pamela/Fermi LAT/Hess favored
region (as in \citealp{Meade:2009iu}) corresponding to the
elliptical spot centered at about $10^3$ GeV and $10^{-23}$
cm$^3$s$^{-1}$.} \label{leptonicbounds}
\end{center}
\end{figure*}

\subsection{Magnetic Field}

Following \citealp{Tabatabaei2008}, the equipartition total magnetic
field strength in the radio cavity is estimated as $7.1\pm0.5\
\mu$G. An exponential decrease is assumed to describe the field
along the $z$ axis:
\begin{equation}
B=B_0\,e^{-|z|/z_{0}},
\end{equation}
where $z_0$ is the scale height of the magnetic field. Observations of
edge-on galaxies give a scale height of about 1.8\,kpc for the
synchrotron emission \citep[see ][]{Beck2009}.  This leads to a
magnetic field scale height of $z_{0}\sim\,7$ kpc, assuming a
nonthermal spectral index of $\sim\,1$ \citep[see e.g.][]{Klein1982}.

In the case in which the equipartition hypothesis were not valid
the situation would be more involved and two cases can be taken into account:

(a) If the field strength is constant across the cavity, the
cosmic ray electrons (CRE) have to be deficient to explain the low
radio emission. $B$ can hardly be constant with height, too, but
it could decrease even slower than with a scaleheight of 7 kpc if
CREs suffer from strong energy losses (which is probable). We
estimated a 7.5 $\mu$G field strength in this case and a
scaleheight that could vary from 5 to 10 kpc;

(b) If however the CRE density is about constant across the
cavity, the field there would be smaller than 7 $\mu$G and, if it
is also constant with height, the $B$--scaleheight would be 2
times smaller. We estimated $B\sim3.2\,\mu$G and $z_0=3.5$\,kpc.
This is however quite unlikely because CREs loose energy with
height.

\subsection{Interstellar Radiation Field}

Once the high electrons are injected, they lose energy due,
mainly, to two physical process: the synchrotron radiation we are
interested in, and inverse Compton scattering off of the ISRF
photons. Therefore, we need to characterize the ISRF in the cavity
to properly subtract the amount of energy lost in the scattering
with the photons.

According to \citealp{Deul1989} we assume for the ISRF in the disk
of M33 $ U_{rad}^{(D)}(R)=5.32\,e^{-R/2.10}\,\chi , $ where
$\chi=0.539$ eV\,cm$^{-3}$ is the MW's ISRF at the Solar System
position \cite{Weingartner2001}. The previous expression is
obtained by setting the ISRF equal to $\chi$ to a galactocentric
radius of $3.5$ kpc. To deduce the scale height of the exponential
decrease of the ISRF far from the disk, we rescale the MW field in
the same way. A galactocentric distance of $2.11$ kpc in M33
corresponds to $5.12$ kpc in the MW. Using the model described in
\citealp{Porter2005} it is possible to deduce that the MW's ISRF
scale height at a radius of $5.12$ kpc is equal to $3.72$ kpc
which, when rescaled to M33, gives $1.53$ kpc.

Putting things together we parameterize the ISRF in the {\it Radio
Cavity} as:
\[
U_{rad}(z)=U_0 \, e^{-|z|/h_0}\ ,
\]
with $U_0=1.05$ eV cm$^{-3}$, and $h_0=1.53$ kpc.

\section{Results and Conclusions}
\label{Sec:conclusions}

\begin{figure*}
\begin{center}
\includegraphics[width=0.48\textwidth]{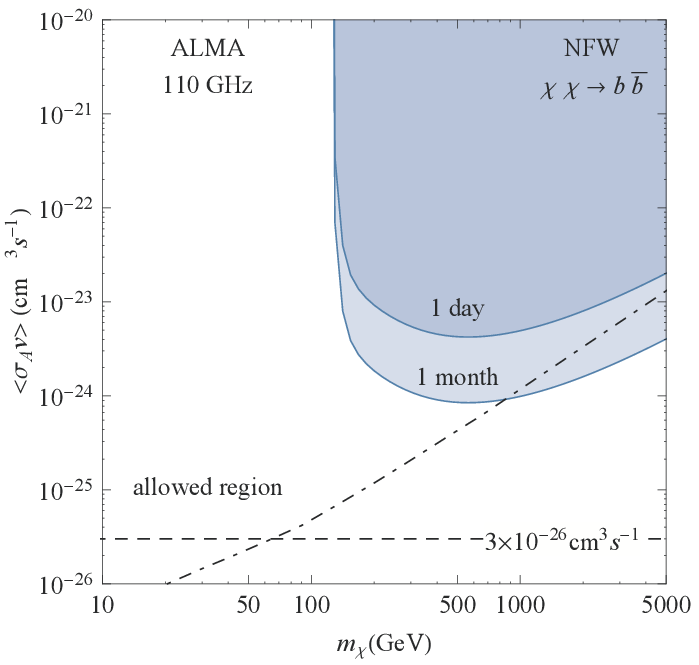}
\includegraphics[width=0.48\textwidth]{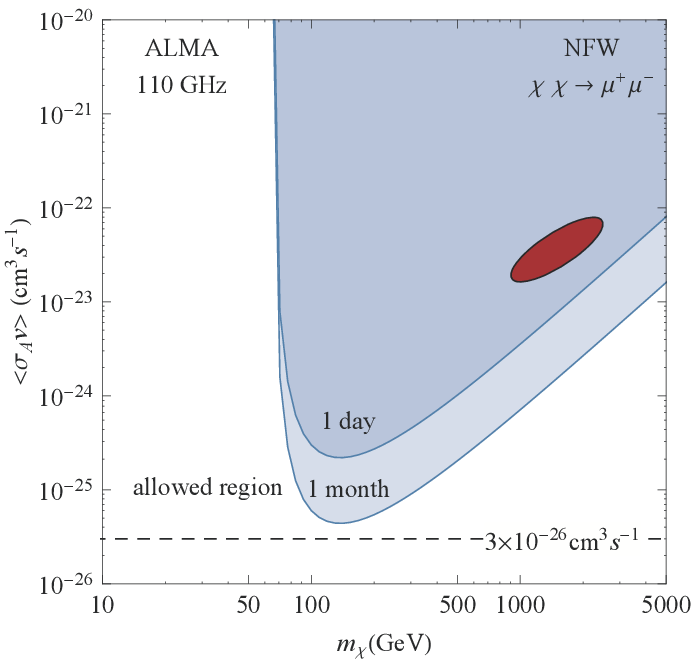}
\caption{Exclusion plots corresponding to a null detection (within
the experimental sensitivity) radio flux. The left panel refers to
DM particles annihilating into $b\,\bar{b}$ quarks derived for one
day and one month of observation. The dot-dashed line represents
the forecast (at 5$\sigma$) for five years of Fermi LAT data
taking obtained by means of Galprop for optimized parameters (see
\citealp{Baltz:2008wd} for details). The right panel refers
instead to annihilation into $\mu^+\mu^-$. The exclusion regions
are superimposed to the ellipse favored by Pamela/Fermi LAT/Hess
data (see \citealp{Meade:2009iu}).} \label{forecasts}
\end{center}
\end{figure*}

Following the approach described above one can compute the
expected radio emission due to electrons/positrons produced in the
annihilation in pairs of WIMPs forming the halo of M33, and in
particular the signal coming from the {\it Radio Cavity} which
cannot overcome the observed signal. Hence, by using the
observations of M33 at $\lambda=3.6$ cm and $\lambda=20$ cm, we
obtained the exclusion plots reported in Fig. \ref{hadronicbounds}
and \ref{leptonicbounds}. Three cases are plotted for each
wavelength and DM density profile: The equipartition hypothesis is
verified (case 1: $B=7.1\,\mu$G, $z_0=7.0\,$kpc); The
equipartition hypothesis is not verified and the field strength is
constant across the cavity (case 2a: $B=7.5\,\mu$G, $z_0=10\,$kpc)
or the CRE density is constant (case 2b: $B=3.2\,\mu$G,
$z_0=3.5\,$kpc). Let us first consider the results shown in Fig.
\ref{hadronicbounds}. For both wavelengths the bounds are compared
to those corresponding to the MW (dot--dashed lines), derived in
the same particle physics hypothesis \cite{Borriello:2009gy} in
which DM particles fully annihilate into $b\bar{b}$ couples. The
results are quite dependent on the DM profile chosen, due to the
position of the radio cavity (quite near the galactic center). In
fact, assuming a Burkert density profile one gets less
constraining results. In fact, at $2.1$ kpc from the galaxy center
the Burkert profile is about $2/5$ of the NFW one, therefore we
obtain six times weaker bounds from the cored case.

It is quite encouraging to observe that, at least for a favored
case of an equipartition magnetic field and for a NFW profile,
this calculation provides comparable or stringent bounds on DM
than the ones obtained from our Galaxy. Moreover the M33 radio
data used were not tailored for this aim, and thus could be
largely improved in terms of resolution (see below).

Let us consider the case of very heavy WIMP directly annihilating
into charged leptons as recently proposed in order to explain the
positron excess observed by Pamela satellite \cite{Adriani:2009}
and the $e^+$-$e^-$ excess seen by Fermi LAT
\cite{Collaboration:2009zk}. This unexpected flux is made of
electrons with energy greater that $\sim 100$ GeV. The synchrotron
emission due to electrons of this energy is peaked around 100 GHz,
quite far from the values we are considering here. Fig.
\ref{leptonicbounds} shows the bounds for the case in which the DM
particle fully annihilate into $\mu^+\mu^-$ couples. In this case
the DM contributes only for $\sim$ 1\,\% to the observed flux at
20 cm, while the situation improves at 3.6 cm, where the DM
accounts for a order 10\,\% of the emission. Therefore, higher
frequency observations would be needed to adequately consider this
case.

Among ground-based telescopes, highest resolution (0.03$\arcsec$)
and sensitivity (0.060 mJy) observations at high-frequencies ($\nu
\sim$ 100~GHz) will be achieved by the Atacama Large Millimeter
Array (ALMA). In addition Rotation Measurements Synthesis
\cite{Heald2000} at as many wave wavelengths as possible will lead
to a more precise magnetic field study, hopefully removing the
degeneracy of possibilities shown in Fig. \ref{hadronicbounds} and
\ref{leptonicbounds}. Such a high angular resolution could improve
our results. In fact if one can identify a small region inside the
cavity from which one detects a vanishing flux (within the
experimental sensitivity) this naturally results in a significant
improvement of our bounds. To be quantitative let us consider ALMA
observing at a 110 GHz in a compact configuration (50 antennas,
3.7 arcsec beam, 16 GHz bandwith). By using the
\citealp{ALMA:calculator} one gets a 0.01 mK surface brightness
sensitivity for one day of exposure time.

In Fig. \ref{forecasts} we report the exclusion plots
corresponding to the above null detection assumption for a NFW DM
density profile. The left panel shows the results obtained for
annihilation into $b\,\bar{b}$ while the right one refers to
$\mu^+\mu^-$. For the hadronic case we compare our bounds, derived
for one day and one month of observation, with the forecast (at
5$\sigma$) corresponding to five years of Fermi LAT data taking
\citealp{Baltz:2008wd}. Interestingly radio observations perform
better for heavy WIMPs. In the right panel our exclusion regions
are superimposed to the ellipse favored by Pamela/Fermi LAT/Hess
data (see \citealp{Meade:2009iu} and references therein).

To summarize, we derived the expected secondary radiation due to
synchrotron emission from high energy electrons/positrons produced
in DM annihilations in M33. The comparison of the expected DM
induced emission with radio continuum observations, especially
focussed on a particular low emitting {\it Radio Cavity}, allows
to obtain bounds in the $m_\chi$--$\sigva$ plane. By using not
tailored archival data we are able to put bounds comparable with
ones obtainable by full sky observations of the radio emission
from our Galaxy. This suggests the possibility that focused
observations of small and well known astrophysical object could
gain a great predictive power. To be more quantitative on the
potentiality of the method we have considered an optimistic
scenario were a high resolution radio telescope like ALMA does not
detect any flux (within the experimental sensitivity) in a region
inside the radio cavity. In this case the bounds we can derive on
WIMPs parameters are better than the corresponding Fermi LAT
forecast in the case of heavy WIMPs decaying into hadrons.
Assuming instead annihilation into muons we could be able to rule
out the region favored by Pamela/Fermi LAT/Hess in a single day of
observation.

\section*{Acknowledgments}
We thank A. Cuoco for useful comments. G.L. acknowledges the
support from the Ministry of Foreign Affairs through an Italy-USA
Bilateral Great Relevance (BUilding an e platform for Data
Mining). Use of the publicly available HEALPix software is
acknowledged. G.M. acknowledges supports by the grant INFN I.S.
Fa51.


\begin{thebibliography}{00}

\bibitem[Abdo et al. 2008]{Collaboration:2009zk}
  Abdo, A. A. et al. 2009, Phys. Rev. Lett., 102, 181101

\bibitem[Adriani et al. 2009]{Adriani:2009}
Adriani, O. et al. 2009, Phys. Rev. Lett. 102, 051101

\bibitem[ALMA Sensitivity Calculator]{ALMA:calculator}
ALMA Sensitivity Calculator,\\
http://www.eso.org/sci/facilities/alma/observing/tools/etc/

\bibitem[Aloisio et al. 2004]{Aloisio:2004hy}
  Aloisio, R., Blasi, P., \& Olinto, A.V. 2004, JCAP, 0405, 007

\bibitem[Baltz \& Edsjo 1998]{Baltz:1998xv}
  Baltz, E. A., \& Edsjo, J. 1999, Phys. Rev.  D, 59, 023511

\bibitem[Baltz \& Wai 2004]{Baltz:2004bb}
Baltz, E. A., \& Wai, L. 2004, Phys. Rev.  D, 70, 023512

\bibitem[Baltz et al. 2008]{Baltz:2008wd}
  Baltz, E. A. et al. 2008, JCAP, 0807, 013

\bibitem[Beck 2009]{Beck2009}
Beck, R. 2009, Ap\&SS, 320, 77

\bibitem[Bergstrom 2000]{Bergstrom:2000pn}
  Bergstrom, L. 2000, Rept. Prog. Phys., 63, 793


\bibitem[Bergstrom et al. 2006]{Bergstrom:2006ny}
  Bergstrom, L., Fairbairn, M., \& Pieri, L. 2006, Phys. Rev. D, 74, 123515

\bibitem[Bertone et al. 2005]{Bertone:2004pz}
  Bertone, G., Hooper, D., \& Silk, J. 2005, Phys. Rept., 405, 279

\bibitem[Blasi et al. 2002]{Blasi:2002ct}
  Blasi, P., Olinto, A. V., \& Tyler, C. 2003, Astropart. Phys., 18, 649

\bibitem[Borriello et al. 2009]{Borriello:2009gy}
  Borriello, E., Cuoco, A., \& Miele, G. 2009, Phys. Rev. D, 79, 023518

\bibitem[Burkert 1995]{Burkert1995}
Burkert, A. 1995, ApJ, 447, L25

\bibitem[Carignan 2006]{Carignan:2006}
Carignan, C., Chemin, L., Huchtmeier, W. K., \& Lockman, F. J. 2006, ApJ, 641, L109

\bibitem[Ciani et al. 1984]{Ciani:1984}
Ciani, A., D'Odorico, S., Benvenuti, P. 1984, A\&A, 137, 223

\bibitem[Cioni et al. 2007]{Cioni:2007}
Cioni, M.-R. L. et al., arXiv:0709.2949

\bibitem[Colafrancesco 2005]{Colafrancesco:2005ji}
  Colafrancesco, S., Profumo, S., \& Ullio, P. 2006, A\&A, 455, 21

\bibitem[Colafrancesco 2006]{Colafrancesco:2006he}
  Colafrancesco, S., Profumo, S., \& Ullio, P. 2007, Phys. Rev. D, 75, 023513

\bibitem[Corbelli 2003]{Corbelli2003}
Corbelli, E. 2003, MNRAS, 342, 199


\bibitem[de Oliveira Costa et al. 2008]{deOliveiraCosta:2008pb}
  de Oliveira-Costa, A., Tegmark, M., Gaensler, B. M., Jonas, J., Landecker, T. L., \& Reich, P.  2008, MNRAS, 388, 247

\bibitem[Deul 1989]{Deul1989} Deul, E. R. 1989, A\&A, 218, 78

\bibitem[Donato et al. 2003]{Donato:2003xg}
  Donato, F., Fornengo, N., Maurin, D., Salati P., \& Taillet, R. 2004, Phys. Rev. D, 69, 063501

\bibitem[Finkbeiner 2004]{Finkbeiner:2004us}
 Finkbeiner,  D. P. 2004, arXiv:astro-ph/0409027

\bibitem[Freedman et al. 1991]{Freedman1991}
Freedman, W. L., Wilson, C. D., \& Madore, B. F. 1991, ApJ, 372, 455.

\bibitem[Gold et al. 2008]{Gold:2008kp}
  Gold, B., et al. 2009, ApJS, 180, 265

\bibitem[Gordon et al. 1999]{Gordon:1999}
Gordon, K. D., Hanson, M. M., Clayton, G. C., Rieke, G. H., \& Misselt, K. A. 1999, ApJ, 519, 165

\bibitem[Haberl \& Pietsch 2001]{Haberl:2001}
Haberl, F.,  \& Pietsch, W.  2001, A\&A, 373, 438

\bibitem[Heald et al. 2000]{Heald2000}
(Heald, G. et al. 2000, A\&A 503, 409)

\bibitem[Hooper \& Silk 2004]{Hooper:2004bq}
  Hooper, D., \& Silk, J. 2005, Phys. Rev. D, 71, 083503

\bibitem[Hooper at al. 2007]{Hooper:2007kb}
  Hooper, D., Finkbeiner, D. P., \& Dobler, G., 2007, Phys. Rev. D, 76, 083012

\bibitem[Jeltema \& Profumo 2008]{Jeltema:2008ax}
  Jeltema, T. E., \& Profumo, S. 2008, ApJ, 686, 1045

\bibitem[Jungman et al. 1995]{Jungman:1995df}
  Jungman, G., Kamionkowski, M., \& Griest, K. 1996, Phys. Rept.,  267, 195

\bibitem[Klein et al. 1982]{Klein1982}
Klein, U., Beck, R., Buczilowski, U. R., \& Wielebinski, R. 1982, A\&A, 108, 176

\bibitem[Komatsu et al. 2008]{Komatsu:2008hk}
  Komatsu, E., et al. 2009, APJS, 180, 330

\bibitem[Massey et al. 2006]{Massey:2006}
Massey, P., Olsen, K. A. G., Hodge, P. W.,  Strong, S. B., Jacoby, G. H.,
Schlingman, W., \& Smith, R. C. 2006, AJ, 131, 2478

\bibitem[Meade et al. 2009]{Meade:2009iu}
Meade, P., Papucci, M., Strumia, A., \& Volansky, T. 2009,
  arXiv:0905.0480 [hep-ph].

\bibitem[Moore et al. 1999]{Moore1999}
Moore, B., Ghigna, S., Governato, F., Lake, G., Quinn, T. R., Stadel, J. \& Tozzi P. 1999, ApJ., 524, L19

\bibitem[Navarro et al. 1996]{Navarro:1996gj}
  Navarro, J. F., Frenk, C.S., \&  White, S. D. M. 1997, ApJ, 490, 493

\bibitem[Porter \& Strong 2005]{Porter2005}
 Porter, T. A., \& Strong, A. W. 2005, astro-ph/0507119

\bibitem[Regis \& Ullio 2008]{Regis:2008ij}
  Regis, M., \& Ullio, P. 2008, Phys. Rev. D, 78, 043505

\bibitem[Tabatabaei et al. 2007a]{Tabatabaei2007c}
Tabatabaei, F. S., et al. 2007, A\&A, 466, 509

\bibitem[Tabatabaei et al. 2007b]{Tabatabaei2007a}
Tabatabaei, F. S., Krause, M., \& Beck, R. 2007, A\&A, 472, 785

\bibitem[Tabatabaei et al. 2007c]{Tabatabaei2007b}
Tabatabaei, F. S., Beck, R., Kr\"ugel, E., Krause, M., Berkhuijsen, E. M., Gordon, K. D. \&  Menten, K. M. 2007, A\&A, 475, 133

\bibitem[Tabatabaei et al. 2008]{Tabatabaei2008}
Tabatabaei, F. S., Krause, M., Fletcher, A., \& Beck, R. 2008, A\&A, 490, 1005

\bibitem[Tasitsiomi et al. 2003]{Tasitsiomi:2003vw}
  Tasitsiomi, A., Siegal-Gaskins, J. M., \& Olinto, A. V. 2004, Astropart. Phys., 21, 637

\bibitem[Weingartner \& Draine 2001]{Weingartner2001}
Weingartner, J. C., \&  Draine, B. T. 2001, ApJ, 134, 263

\bibitem[Zhang \& Sigl 2008]{Zhang:2008rs}
  Zhang, L., \& Sigl, G. 2008, JCAP 09, 027

\end{thebibliography}
\end{document}